%% file: req.tex
\documentclass[conference]{IEEEtran}
\IEEEoverridecommandlockouts
\usepackage[square,sort,comma,numbers]{natbib}
\usepackage[T1]{fontenc}
\usepackage{upquote}
\usepackage[numbers]{natbib}
\usepackage{amsmath,amssymb,amsfonts}
\usepackage{algorithmic}
\usepackage{graphicx}
\usepackage{textcomp}
\usepackage{comment}
\usepackage{framed}
\usepackage{hyperref}
\usepackage{verbatim}
\usepackage{adjustbox,lipsum}
\usepackage{xcolor}
\def\BibTeX{{\rm B\kern-.05em{\sc i\kern-.025em b}\kern-.08emT\kern-.1667em\lower.7ex\hbox{E}\kern-.125emX}}
\usepackage{comment}
\usepackage{ifthen}
\newboolean{showcomments} 
\setboolean{showcomments}{True}
 
\usepackage{csquotes}
\begin{document}

\title{Identifying human values from goal models: An
industrial case study\\}

\author{
    \IEEEauthorblockN{Tahira Iqbal\IEEEauthorrefmark{1}, Kuldar Taveter\IEEEauthorrefmark{1}, Tarmo Strenze\IEEEauthorrefmark{1}, Waqar Hussain\IEEEauthorrefmark{2},Omar Haggag\IEEEauthorrefmark{3},  John Alphonsus Matthews\IEEEauthorrefmark{1}, Anu Piirisild\IEEEauthorrefmark{1} }
    \IEEEauthorblockA{\IEEEauthorrefmark{1}Institute of Computer Science, University of Tartu,  Estonia
    \\\{tahira.iqbal, kuldar.taveter, tarmo.stremze, john.matthews, anu.piirisild\}@ut.ee}
    \IEEEauthorblockA{\IEEEauthorrefmark{2}Data 61, CSIRO, Australia
    \\\{waqar.hussain\}@data61.csiro.au}
        \IEEEauthorblockA{\IEEEauthorrefmark{3}Faculty of Information Technology, Monash University, Australia
    \\\{omar.haggag\}@monash.edu.au}

}

\maketitle
\begin{abstract}
Human values are principles that guide human actions and behaviour in personal and social life. Ignoring human values during requirements engineering introduces a negative impact on software uptake and continued use. Embedding human values into software is admittedly challenging; however, early elicitation of stakeholder values increases the chances of their inclusion into the developed system. Using Pharaon, a research and innovation project of the European Union’s Horizon 2020 program, as a case study we analysed stakeholder requirements expressed as motivational goal models consisting of functional, quality, and emotional goals in three large-scale trial applications of the project. We were able to elicit 9 of 10 human values according to the theory of human values by Schwartz from the motivational goal models that represent the requirements for the three applications. Our findings highlight the dominant trend of stakeholder values being embedded in emotional goals and show that almost 45\% of the identified values belong to the value categories of Security and Self-direction. Our research extends prior work in emotional goal modelling in requirements engineering by linking emotional goals to various stakeholder roles and identifying their values based on the Schwartz theory of human values.
\end{abstract}

\newcommand{\tablepath}{Tables}
\input{Introduction.tex}

\input{Background}
\input{Study_Design}

\input{Results}

\input{Disscussion.tex}

\input{Relatedwork}
\input{Conclusion.tex}

\bibliographystyle{unsrt}
\small{
    \bibliography{./req}  
}
\end{document}

%% file: Introduction.tex
\section{Introduction}
\label{sec:intro}

Human values express what is important for an individual or a society \cite{schwartz2012overview}. They are the principles that guide human actions and behaviour in daily life \cite{rokeach1973nature}. In recent years, human value research has been gaining attention in software engineering \cite{nurwidyantoro2022human,perera2020study,perera2020continual,ferrario2016values}. The existing studies \cite{wang2013affects,harris2016identifying} have shown the importance of considering human values, e.g. they may significantly influence one's intention to install and use the software. Also, violating or ignoring values can cause adverse outcomes such as damage to a company's reputation, and loss of profits \cite{georgievski2016effect}. For example, the Facebook Cambridge Analytica scandal disclosed a data breach in which users' private data was used for political reasons without their consent. With this, Facebook violated user values, such as Security and Trust. 
As a result, users' trust in Facebook dropped by 66\%,  it suffered significant reputational damage, faced harsh public scrutiny, and experienced a significant decline in profits \cite{bbc}. The existing research mainly focuses on the identification of human values from different sources such as issue trackers \cite{nurwidyantoro2022human}, user stories \cite{hussain2022can}, and app reviews \cite{obie2021first}. However, little attention has been paid to other requirements engineering (RE) artefacts, such as goal models and in particular motivational goal models \cite{sterling2009art,miller2014requirements,miller2015emotion,taveter2019method,Sulis2022}, which constitute an important artefact for representing early requirements \cite{horkoff2016interactive, van2009requirements}.

The methods of conceptual modelling using goal models and value models to develop digital healthcare solutions carry significant limitations. For example, the value models described in \cite{henkel2007value} mainly focus on increasing the business value of the service to patients instead of focusing on patients' emotions and human values.
Similarly, the authors in \cite{vermunt2018three}  developed a three-level goal hierarchy constituting disease-specific goals, functional goals and fundamental goals, representing patients' life priorities. The authors argue that explicating fundamental goals or values can help identify potential mismatches between medical standards and patient values, for example in treating multi-morbidity. 
While useful for conceptualising patient goals, the model ignores the consideration of goals and value-oriented requirements for any other category of stakeholders, such as caregivers or relatives. The work described in \cite{aidemark2021contextual} provides a checklist that includes patient long-term goals that aid in developing healthcare technology to support patients’ recovery process in their home environment. Again, this work only caters for the goals of patients. Recent research on emotion-oriented requirements \cite{curumsing2019emotion} that relates more closely to our work has attempted to link functional and emotional goals with system requirements for a wider group of  stakeholders. However, the study does not apply any existing theory of human values to map the emotional goals to the values of multiple stakeholder groups. 

Motivated by this, to the authors' knowledge, our study is the first kind of study that explores the presence of human values in motivational goal models \cite{sterling2009art,miller2014requirements,miller2015emotion,taveter2019method,Sulis2022}. To address this research gap, we investigate the following research questions:

\begin{itemize}
    \item RQ1: Can we identify human values in motivational goal models? 
    \item RQ2: What are the trends of human values identified in motivational goal models?
\end{itemize}

 The contribution of our work is establishing the link between emotional goals of motivational goal models and human values based on the Schwartz theory. For doing this, we demonstrated the identification of human values in the industrial research and development project Pharaon\footnote{https://www.pharaon.eu/}, which consists of 40 partners. Overall, we identified 9 out of 10 value categories, and 28 out of 58 value items based on the Schwartz theory of human values \cite{schwartz2012overview} from six motivational goal models developed in the Pharaon project. Almost 45\% of the identified values in our project belonged to the value categories of \textit{Security} and \textit{Self-direction}. Our finding is a significant insight that highlights the dominant stakeholder expectations beyond the main functional requirements for the offerings of digital healthcare and well-being services for older adults and the related stakeholder groups.

The rest of this paper is structured as follows. In Section \ref{sec:Background}, the notion of human values and goal-based requirements modelling are explained. In Section \ref{sec:Methodology}, the research methodology is described, consisting of the descriptions of the case study and how the data collection and analysis were performed. In Section \ref{sec:Results}, the results are presented. The results are discussed in Section \ref{sec:Discussion}, where also answers to the research questions are provided and threats to validity are analysed. The related work is overviewed in Section \ref{sec:Related}. Finally, conclusions are drawn in Section \ref{sec:conclusion}.

%% file: Background.tex
\section{Background}
\label{sec:Background}
\input{Tables/humanvalue_items_def}
\subsection{Human Values}

The concept of ``human values'' has been thoroughly researched in psychology and sociology. Notable theories are the value survey by Rokeach \cite{rokeach1973nature}, culture dimensions by Hofstede and Bond \cite{hofstede1984hofstede}, theory of basic human values by Schwartz \cite{schwartz2012overview, schwartz2017refined}, and functional theory of human values by Gouveia and others \cite{gouveia2014functional}. Human values can be defined as principles that guide human actions and behaviour in daily personal and social life \cite{rokeach1973nature} and modes of conduct that a person likes or chooses among different situations \cite{parashar2004perception}. In this study, we use the theory of basic human values by Schwartz \cite{schwartz2012overview} to link emotional goals with the corresponding value categories and value items included in the Schwartz theory of human values. We chose the Schwartz theory over the other theories of human values for three reasons. First, the Schwartz theory has been widely used in software engineering \cite{ferrario2016values, whittle2019case, nurwidyantoro2022human, hussain2022can} and requirements engineering \cite{proynova2011investigating, alatawi2018psychologically, perera2020continual, obie2021first}. Second, the theory itself has been extensively validated in various domains and findings from across different cultures \cite{schwartz2001extending} and in 83 countries \cite{schwartz2012refining}.  Third, according to \cite{cheng2010developing}, the Schwartz theory subsumes other theories of human values.

The Schwartz theory \cite{schwartz2012overview} distinguishes ten categories of human values and divides them into the main categories with the (i) personal focus; and (ii) social focus. Values with a personal focus are linked to either self-enhancement or a person's openness to change. The values with a social focus are concerned with how one relates socially to others and affects their interests. A complete list of these ten categories along with their definitions is provided in Table. I. Also listed in the table are all 52 individual value items identified by Schwartz.



\subsection{Motivational Goal Modelling}


Humans are goal-directed creatures. Human motivation energizes, directs, and sustains their goal-directed activities \cite{schunk2012motivation}. Motivational modelling \cite{miller2012understanding,lorca2018teaching,burrows2019motivational} is a method that allows ethnographers and requirements engineers to elicit and represent emotional requirements for sociotechnical systems \cite{baxter2011socio} related to the goals to be achieved. In motivational modelling, three kinds of goals – \textit{do}, \textit{be}, and \textit{feel} goals – are elicited from project stakeholders. \textit{Do} goals or functional goals describe what the system to be designed should do or achieve, \textit{be} goals or quality goals describe how the system should be, or more specifically the quality characteristics of the things to be done or achieved, and \textit{feel} goals or emotional goals describe how a user performing a particular role should feel when using the system or, in other words, what emotions should be constructed in the brain of the user when he or she uses the system. The difference between \textit{do} and \textit{feel} goals is that while functional goals characterize the system to be designed and created, emotional goals describe what emotions using the system should construct in the minds of its users, based on the theory of constructed emotion \cite{iqbal2022theory,taveter2019method}. The meaning of a functional goal as compared with activity or task in this context is that a goal is not directly executable, i.e., it leaves open \textit{how} it should be done or achieved. 

\begin{figure}[ht]
    \makebox[\linewidth]{\includegraphics[width=1.0\linewidth]{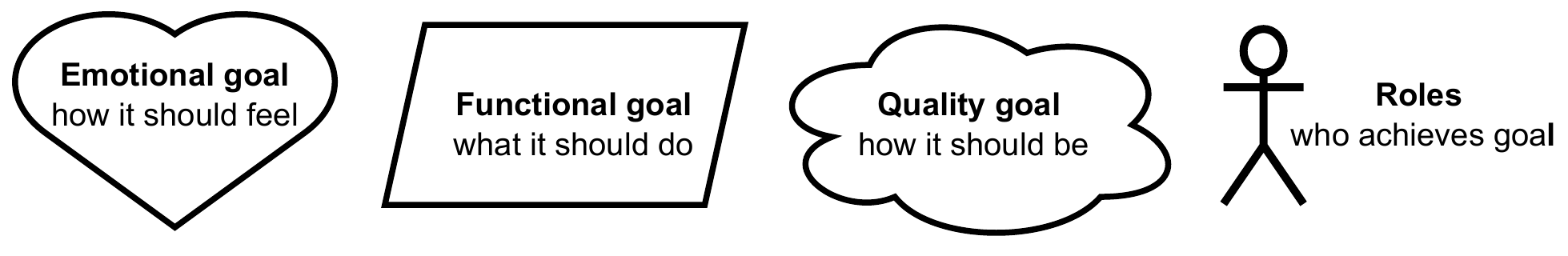}
   }
\caption{Notation for motivational goal models}
\label{Notation}
\vspace{-4mm}
\end{figure}

Motivational goal modelling represents user requirements by hierarchical goal models that decompose a high-level purpose of a system to be designed into lower-level goals, each of which represents a particular aspect of achieving its parent goal \cite{sterling2009art,miller2014requirements,miller2015emotion,taveter2019method,Sulis2022}.  The notation for representing motivational goal models is shown in Figure \ref{Notation}. The skeleton of a goal model is a hierarchy of functional goals drawn as a tree in a top-down manner, starting with the highest-level goal, which represents the purpose of the system. The hierarchical structure is used to indicate that achieving a sub-goal represents an aspect of achieving its parent goal. The roles, quality goals and emotional goals are attached to relevant functional goals at an appropriate level in the goal hierarchy. They also apply to the sub-goals of each relevant functional goal. An emotional goal is also attached to one or more roles, indicating how the performers of the respective roles should feel.

An advantage of motivational goal modelling over other goal modelling methods is that it enables to represent emotional requirements as first-class citizens in the form of emotional goals \cite{taveter2019method}. Motivational goal modelling also supports well the communication between technical and non-technical stakeholders \cite{pedellemotions,wachtler2018development,taveter2019method,mooses2021agent,zimmer2021requirements,moosesmethods,iqbal2022theory,Sulis2022}, which is very relevant in our case study.  

 
According to Schwartz \cite{schwartz2012overview}, value items can be viewed as one's motivational goals. However, values and value items are still more abstract than goals used in motivational goal modelling. Thew and Sutcliffe \cite{thew2018value} emphasize the importance of emotions by the stakeholders in understanding the values embedded in a socio-technical system. Maio \cite{maio2010mental} states that ‘emotions are the primary source of information for consensually important values’ and Schwartz notes that ‘when values are activated, they become infused with feeling' so that people feel aroused when their values are threatened \cite{schwartz2012overview}. In requirements engineering, emotions concerned with human values can be represented as emotional requirements in the form of emotional goals of motivational goal models. However, emotions alone are insufficient for identifying values from motivational goal models. We also need to know what a person performing a particular role should do or achieve. For example, the human values identified for the emotional goal Informed differ for the roles of Older Adult and Informal Caregiver because they need to feel informed about different things.

This paper addresses identifying human values from motivational goal models via emotional goals, for which contextual information is provided by roles and functional goals associated with the emotional goals.

%% file: Tables/humanvalue_items_def.tex
\begin{table*}[ht]
\centering
\footnotesize
\resizebox{\textwidth}{!}{%
\begin{tabular}{|p{4cm}|p{6cm}|p{6cm}|}
\hline
\textbf{Value Category} &
  \textbf{Definition} &
  \textbf{Value Items} \\ \hline
Self-direction &
 Independent thought and action's choosing, creating, exploring &
  Freedom, Creativity, Independence, Privacy, Choosing own goals, Curiosity, Self-respect \\ \hline
Stimulation &
  Excitement, novelty, and challenge in life &
  Excitement in life, A varied life, Daring \\ \hline
Hedonism &
  Pleasure or sensuous gratification for oneself &
  Pleasure, Self-indulgent, Enjoying life \\ \hline
Achievement &
  Personal success through demonstrating competence according to social standards &
  Ambitious, Influential, Capable, Successful, Intelligent \\ \hline
Power &
  Social status and prestige, control or dominance over people and resources &
  Wealth, Authority, Preserving my public image, Recognition, Social power \\ \hline
Security &
Safety, harmony, and stability of society, of relationships and of self &
  National security, Family security, Sense of belonging, Healthy, Clean, Reciprocation of favors \\ \hline
Conformity &
  Restraint of actions, inclinations, and impulses likely to upset or harm others and violate social expectations or norms &
  Obedient, Self-discipline, Politeness, Honoring of parents and elders \\ \hline
Tradition &
  Respect, commitment, and acceptance of the customs and ideas that one’s culture or religion provides &
  Respect for tradition, Devout, Detachment, Humble, Moderate, Accepting my portion in life \\ \hline
Benevolence &
  Preserving and enhancing the welfare of those with whom one is in frequent personal contact &
  Helpful, Responsible, Forgiving, Honest, Loyal, A spiritual life, True friendship, Meaning in life, Mature love \\ \hline
Universalism &
 Understanding, appreciation, tolerance, and protection for the welfare of all people and for nature &
  Equality, Wisdom, Inner harmony, A world of beauty, Social justice, Broadminded, A world at peace, Unity with nature, Protecting environment \\ \hline
  
\end{tabular}%
}
\label{tab:value_deinition}
\caption{Value categories, descriptions, and value items \cite{schwartz2012overview}}
\end{table*}

%% file: Study_Design.tex
\section{Methodology}
\label{sec:Methodology}
\begin{figure*}[ht]
\centering
    \adjincludegraphics[height=5cm,trim={0 {.45\height} 0 0},clip]{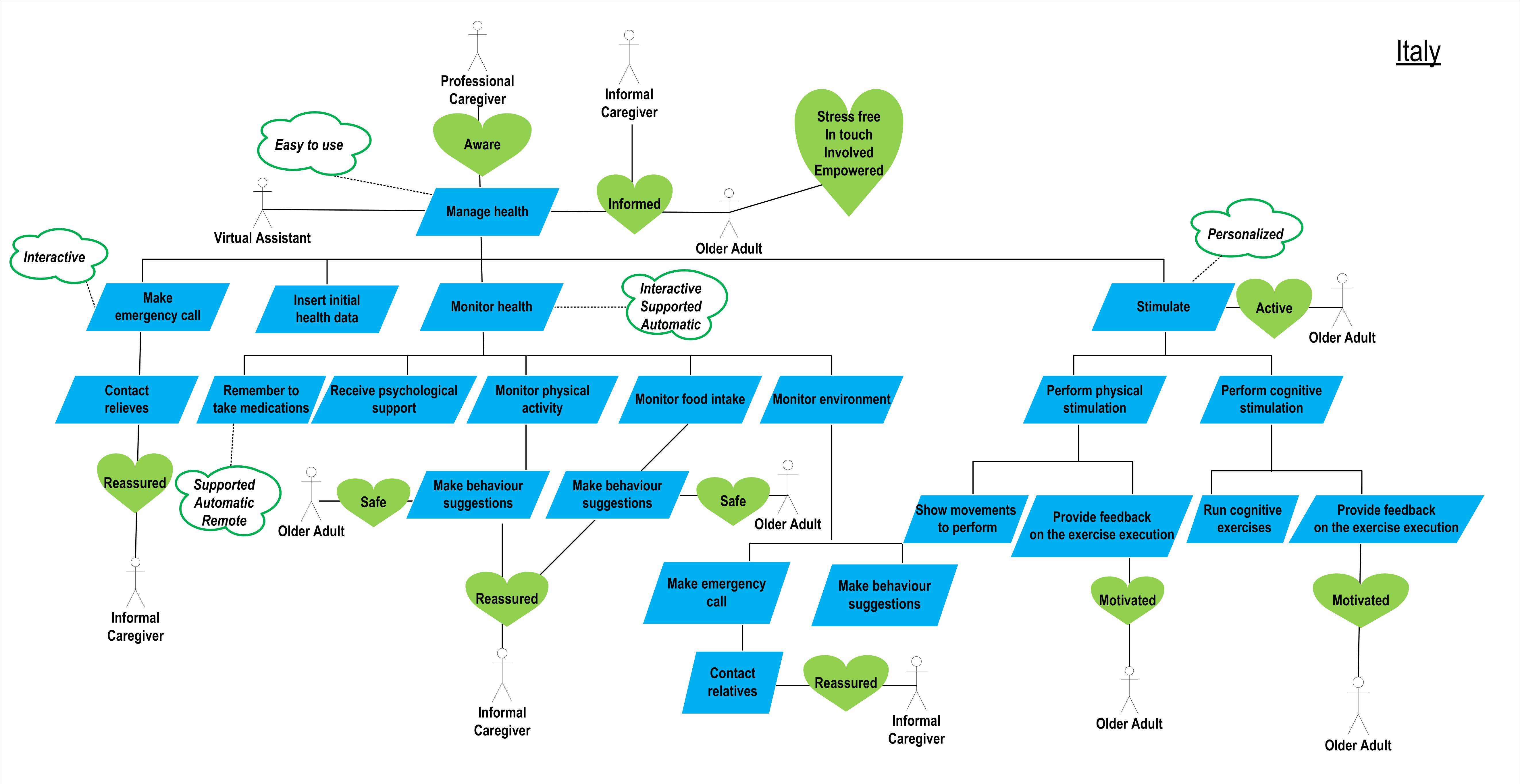}
\caption{Goal model from the Italian pilot}
\label{fig:italian_goal_model}
\vspace{-4mm}
\end{figure*}

In this section, we first briefly discuss the Pharaon project and the data collection process. Furthermore, we describe the manual analysis of motivational goal models that we performed to identify human values in goal models. We also discuss the protocol and steps followed in our analysis.
\subsection{Case Study}
Pharaon is a project within the EU Horizon 2020 research and innovation funding program (EU Grant Agreement 857188).
The purpose of the Pharaon project is to facilitate older adults to be independent and healthy and for their caregivers to have a comfortable environment with up-to-date information and communication technology solutions for taking good care of their relatives and patients. This is achieved by a set of integrated, highly customizable, and interoperable open platforms with advanced services, devices, and tools.
Pharaon has 40 partners  from 12 European countries, contributing to the  development of open platforms and testing several digital solutions in six different pilots carried out in the following five European countries: Italy (the pilot sites of Tuscany and Apulia), Spain (the pilots of Murcia and Andalusia), the Netherlands, Slovenia, and Portugal (the pilot sites of Coimbra and Amadora). Pharon has 15 industrial partners, 4 research organisations, 7 universities, 3 healthcare providers, 3 public authorities, 6 non-governmental organisations, and 2 standardisation bodies. The project lasts from the beginning of December 2019 until the end of July 2024.

A user-centric approach is used in the Pharaon project to maximize the final usability and acceptance of the open socio-technical ecosystem being created by all stakeholders. The main end users of the Pharaon ecosystem will be older adults, whereas several other stakeholders, such as healthcare professionals, formal and informal caregivers, volunteers, governmental and non-governmental organizations, and others, have been identified and involved. The input gathered during the requirements elicitation process was utilized for representing user and pilot requirements for all six pilot sites from five countries involved in the project. The requirements were elicited by different means: face-to-face and virtual stakeholder workshops, online interviews and questionnaires, phone interviews, and a review of the literature including previous projects \cite{moosesmethods}. The project partners of the respective pilots represented early requirements as motivational goal models and transformed them later on into scenarios of motivational modelling \cite{sterling2009art} and into user stories \cite{cohn2004user,oliveira2021transitioning,tenso2017enhancing}. The requirements served as a foundation to define the initial architecture of the Pharaon sociotechnical ecosystem and will guide the project activities to achieve and deploy a successful final system. For the focused and detailed analysis presented in this paper, we have randomly chosen three pilots: Italy, the Netherlands, and Slovenia.  


\subsection{Data Collection}
\label{sec:Data Collection}
We manually analyzed the goal models from three pilots\footnote{The term \textit{pilot in our paper refers to the trial application partners}} of the Pharaon project to discover human values. In the analysis presented in this paper, we included six goal models from three pilots. Fig. \ref{fig:italian_goal_model} shows the goal model of the Italian pilot\footnote{The other five goal models can be accessed online: https://doi.org/10.6084/m9.figshare.21583404.v1}. The notation for motivational goal models is shown in Fig. \ref{Notation}. According to the goal model represented in Fig. \ref{fig:italian_goal_model}, the highest-level goal -- or purpose -- of the sociotechnical system modelled in the figure is to Manage health.
\begin{figure*}[ht]
\centering
\includegraphics[scale= 0.85]{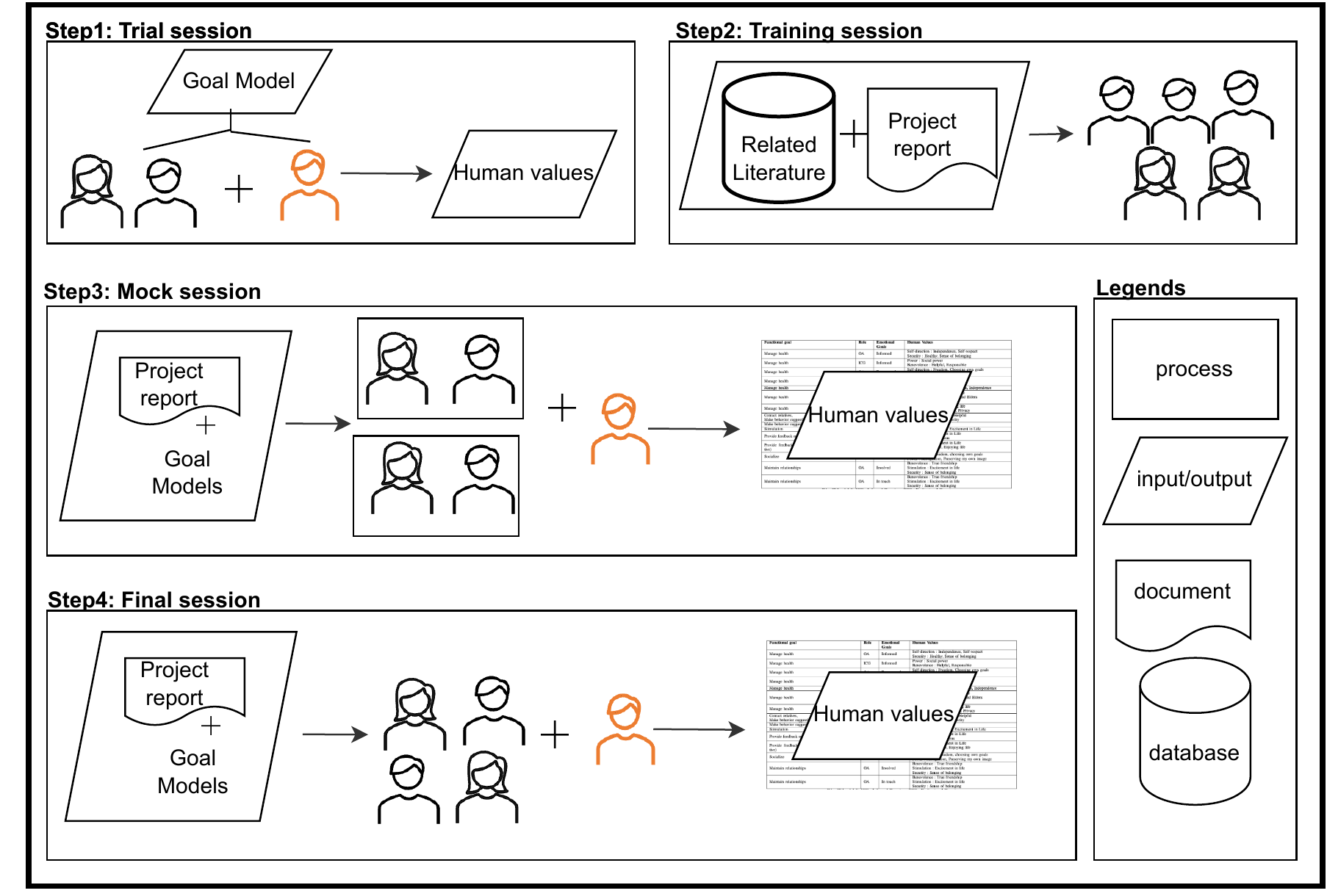}
\caption{The process of mapping human values}
\label{process}
\vspace{-6mm}
\end{figure*}
The main stakeholder roles are Older Adult (whose health is managed), Informal Caregiver and Professional Caregiver. Roles are attached to the functional goals for the achievement of which they are responsible. Functional goals can be associated with emotional goals that model how the stakeholders performing the corresponding roles should feel when achieving the functional goals. Examples of emotional goals when achieving the functional goal Manage health are Empowered for Older Adult, Informed for Informal Caregiver, and Aware for Formal Caregiver. Quality goals like Easy to use represent quality aspects of how the corresponding functional goals, like Manage health in the given case, should be achieved. The functional goal Manage health has been elaborated into four second-level functional goals. Each of these functional goals represents a particular aspect of achieving its parent goal, whereas the order of achieving the functional goals is not represented and will be defined only by scenarios of motivational modelling \cite{sterling2009art}. Some of these functional goals are elaborated further, following the same principle. Thus, roles, emotional goals and quality goals have been attached to the corresponding functional goals, when necessary.

\subsection{Data Analysis}
\label{sec:Data Analysis}
Our study involved five analysts (authors of the paper) ranging from PhD students to associate professors. To avoid biasedness, our analysts consisted of 2 females (40\%), and 3 males (60\%). 

The overall analysis process of mapping human values from emotional goals in the context of the respective roles and functional goals is represented in Fig. \ref{process}. Before initiating the analysis process, we ran a trial session to learn if it was feasible to map emotional goals in the context of roles and functional goals to human values and if the outcome was reasonable. For this, we analyzed the goal model representing the requirements of the Andalusian pilot in Spain. Two analysts (the first two authors of this paper) who have software engineering background and sufficient knowledge about the Schwartz theory of human values individually assigned human values to emotional goals, considering also their contextual roles and functional goals. Later, they discussed the results with the social scientist who is an expert in the Schwartz theory (illustrated in orange color in Fig. \ref{process}). We concluded from the trial session, that the identification of human values from emotional goals was possible, and the results were promising. Therefore, we planned the data analysis with five analysts as the next step. Please note that the results from the above-described trial session are not included in any further data analysis.

In the training session, the analysts attended a joint workshop where Pharon was explained and also we demonstrated the goal model that was used in the trial session and discussed the mapping process. Each analyst was provided with supporting materials, including the project report on the requirements, goal models, and literature on human values. As a result, all analysts had a reasonable understanding of the project and human values before starting the analysis part.  

As human values are subjective, we performed a mock session with all five analysts to bring everyone to the same level of interpretation and understanding of human values, Pharaon project, and motivational goal models. We divided the analysts into two groups, each consisting of two members. Each group individually analyzed one motivational goal model and assigned human values to emotional goals. The remaining analyst, whose background is in social science, moderated the discussion session as an expert in the Schwartz theory of human values. In the discussion session, each group shared their results with the expert and the second group. All the results were discussed, and the expert analyst finalized human values and value items for emotional goals with a mutual agreement. This mock session mainly aimed to understand the process and criteria employed by the analysts to identify human values from emotional goals in the context of roles and functional goals. 

After finishing the mock session, we provided each of the five analysts with all of the goal models from the three pilots to assign human values to emotional goals. At this point, each member was familiar with the process and confident to perform this kind of mapping because of the mock session conducted earlier. In this phase, each analyst analysed goal models for value items associated with human values by Schwartz. While performing the analysis, each analyst also considered project-related information in addition to goal models. The analyst was allowed to propose any human value from the Schwartz value theory.

After finishing the individual analysis, we conducted a joint workshop session for each pilot with all five analysts. The workshop was moderated by the social scientist, and followed a negotiated agreement method \cite{campbell2013coding,morrissey1974sources} to resolve any disagreements and conflicts. Using the negotiated agreement method, all analysts collaboratively agreed on the label of an item under review. In our analysis, the label was a human value attached to an emotional goal. According to the source \cite{krishtul2022human}, this approach is particularly useful
for addressing reliability issues of codes when there are multiple
categories as opposed to two categories where a Cohen’s Kappa
measure would suffice. We resolved conflicts through discussion and finalized the human value categories and value items with mutual agreement. As an output of the analysis, we created tables containing emotional goals mapped to the corresponding human value categories and value items. The tables also included functional goals and roles associated with emotional goals.


\begin{figure*}[ht]
\centering
\includegraphics[scale=0.65]{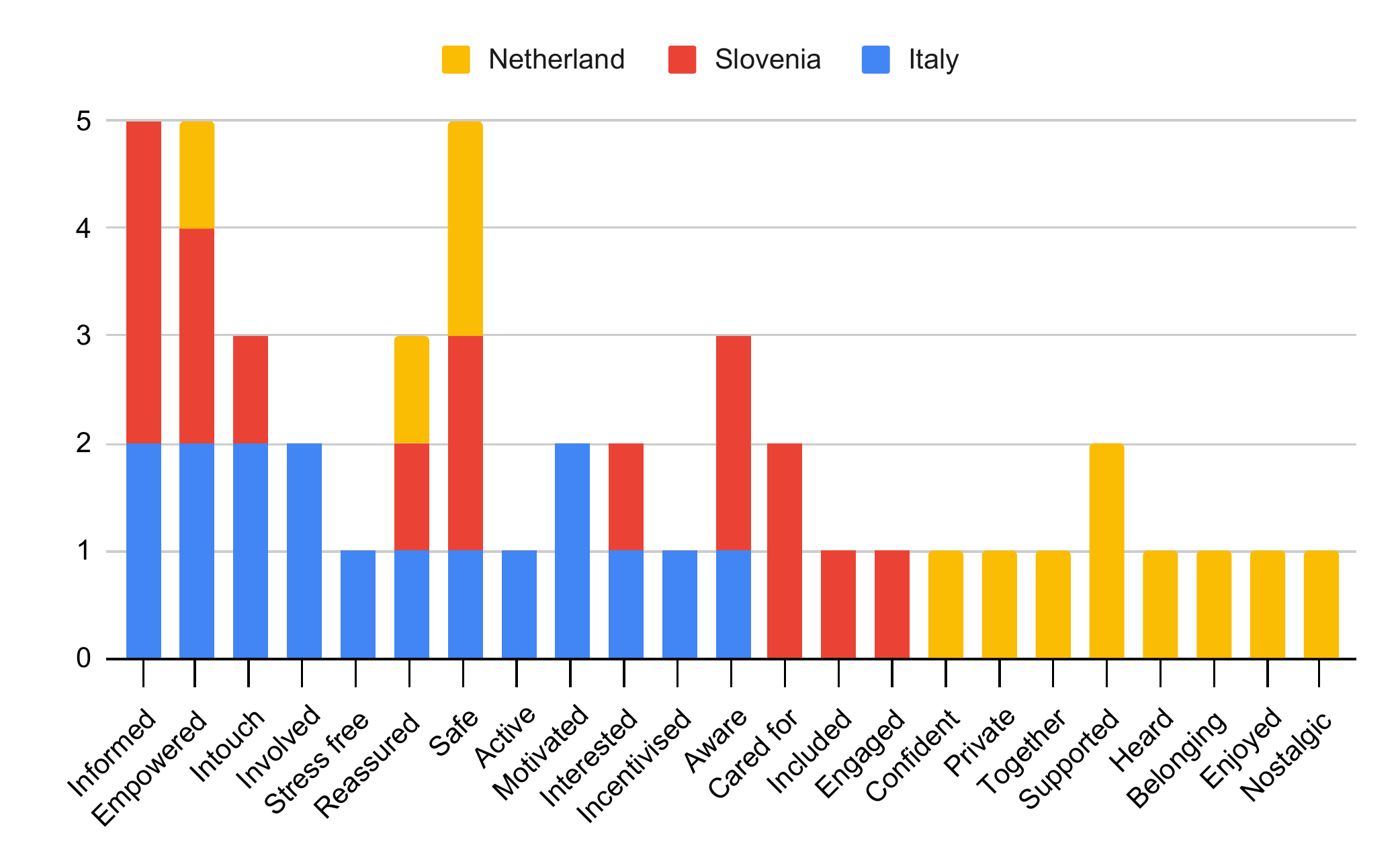}
\caption{Frequencies of emotional goals for each pilot}
\label{Frequencies_of_emotional goal_ per_pilot}
\vspace{-6mm}
\end{figure*}

\textbf{Qualitative Analysis of Goal Models}.
\label{sec:Qaulaitiative Analysis}
The overall qualitative analysis of all emotional goals from the project is illustrated in Fig. \ref{Frequencies_of_emotional goal_ per_pilot}, which shows the frequency of each emotional goal. We retrieved 46 emotional goals from six goal models that were developed by three pilots: Italy, Slovenia, and the Netherlands. Out of 46 emotional goals, 23 were distinct emotional goals. The emotional goals with the highest occurrence frequency were Informed, Empowered, and Safe. Emotional goals with the second highest occurrence frequency were In touch, Reassured, and Aware. The maximum and minimum number of occurrences for emotional goals were respectively 5 and 1.
As is shown in Fig. \ref{Frequencies_of_emotional goal_ per_pilot}, Italian pilot shared more similar emotional goals with the Slovenian pilot (7 goals) compared with the Dutch pilot (3 goals). This can be explained by some similarities between the Italian and Slovenian pilots. The highest number of distinct emotional goals originated in the Dutch (8 goals) and Italian (5 goals) pilots. However, the Slovenian pilot only had three distinct emotional goals.

%% file: Results.tex
\section{Results}
\label{sec:Results}
\input{Tables/Italian_human_value}
To answer RQ1, which focuses on the identification of human values in goal-based requirements, we qualitatively analyzed six goal models with all the emotional goals from three pilots, Italy, Slovenia, and the Netherlands, as is described in Section \ref{sec:Data Collection}. After finishing the analysis phase, we created one table for each pilot. A row of each table consists of a functional goal, role, an emotional goal and identified human value. For example, Table \ref{tab:italian_human_value} was created for the Italian pilot after concluding the analysis process for the goal model shown in Fig. \ref{fig:italian_goal_model}, together with another goal model from the Italian pilot\footnote{Due to space limitations, the other two tables are available online at: \url{https://doi.org/10.6084/m9.figshare.21583404.v1}}. 

In the analysed three pilots of the Pharaon project, we were able to assign human values to emotional goals in the context of the respective roles and functional goals. The steps that we took for the mapping were as follows: (i) running a trial session with two analysts to establish the feasibility of mapping human values from motivational goal models; (ii) running a mock session of mapping emotional goals to human values initially with the five analysts in two groups, each consisting of two members, and thereafter in a joint discussion session, which was moderated by a social scientist who is an expert in the Schwartz theory of human values; (iii) assigning by each of the five analysts individually human values to emotional goals included by the motivational goal models of the three pilots; (iv) conducting for each pilot a joint workshop session with all of the five analysts, moderated by the expert in the Schwartz theory, and discussing the obtained results, where conflicts were resolved through discussion and the human values and value items mapped from emotional goals were finalized with a mutual agreement. Therefore, we answer RQ1 by stating that it is possible to identify human values in motivational goal models.

\begin{figure}
    \centering
    \includegraphics[scale=0.45]{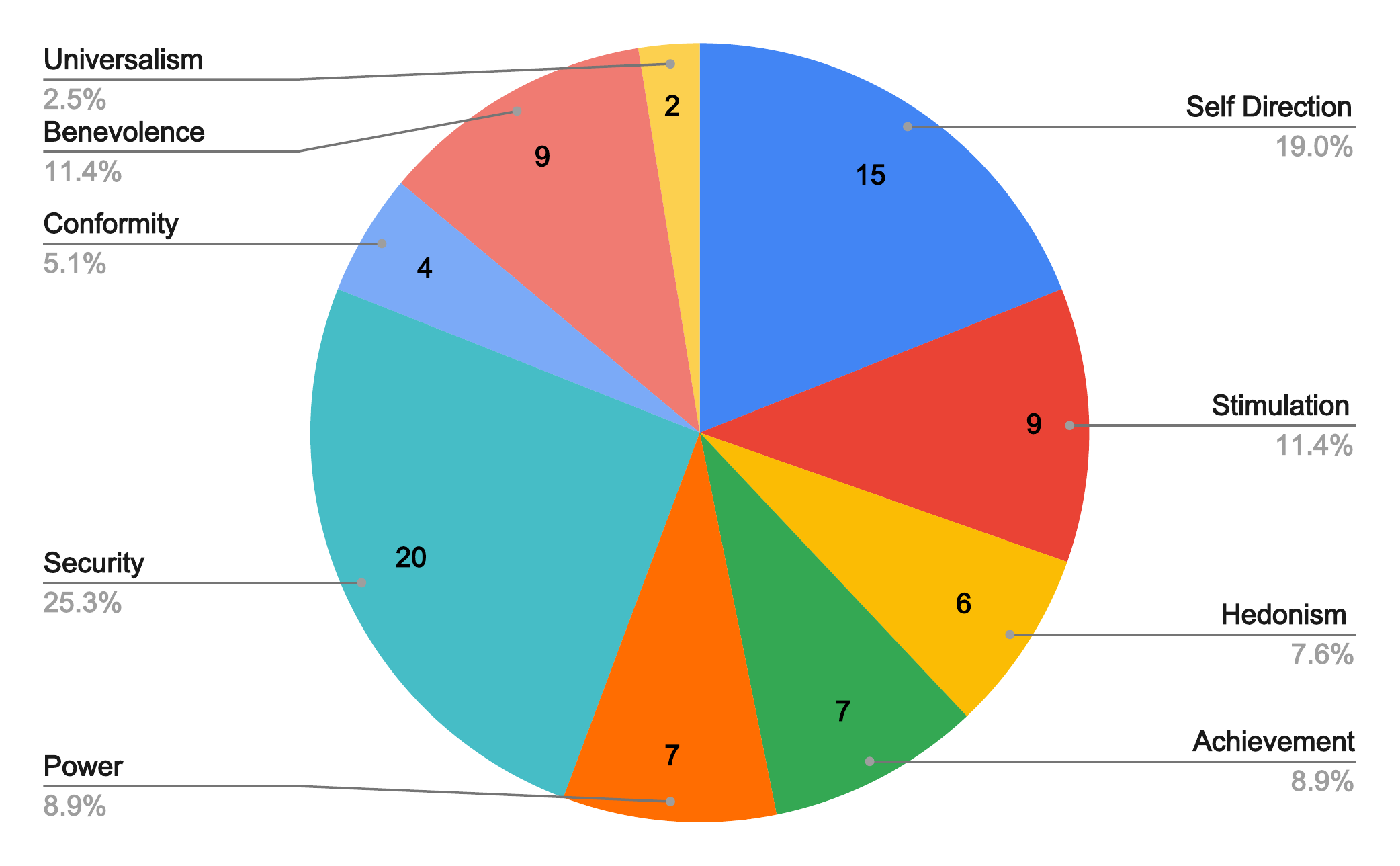}
    \caption{Human values in the pilots}
    \label{fig:values}
    \vspace{-6mm}
\end{figure}

\begin{figure*}
    \centering
    \includegraphics[width=\linewidth]{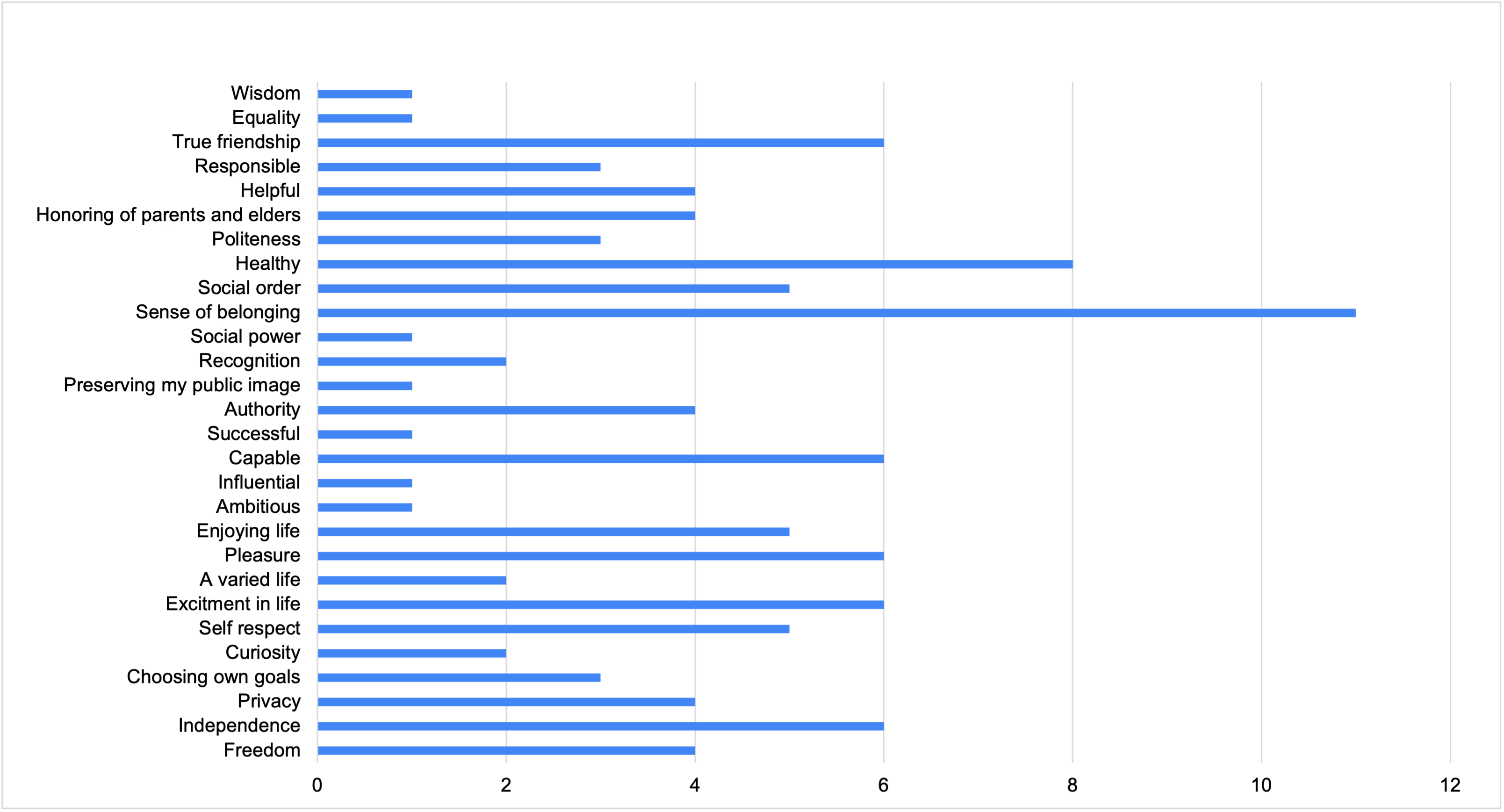}
    
    \caption{Presence of value items}%
    \label{fig:items}
    \vspace{-6mm}
\end{figure*}

Fig. \ref{fig:values} answers RQ2, which is dealing with the trends of human values identified in the goal models, by showing the total count of human values present in the three pilots. The figure contains the main value categories Self-direction, Stimulation, Hedonism, Achievement, Power, Security, Conformity, Benevolence, and Universalism. We identified 9 out of 10 human values by Schwartz in the pilots, with the total occurrence count of 79 times. Our analysis did not find any occurrence of the human value Tradition. The majority of the emotional goals refer to the human values Security and Self-direction, with the respective occurrence frequencies 20 and 15. The least prominent values appeared to be Universalism and Conformity, with the respective occurrence frequencies 2 and 4. Furthermore, we identified 28 out of 52 value items from the Schwartz theory of human values, as is shown in Fig. \ref{fig:items}. The value items Sense of belonging and Healthy from the value category Security rank as the top value items identified in the project. The value item Sense of belonging was mapped from a maximum number of emotional goals, including Informed, In touch, Supported, Cared for, Belonging, Included, and Involved.
\begin{figure}
 \centering
    \includegraphics[scale= 0.5]{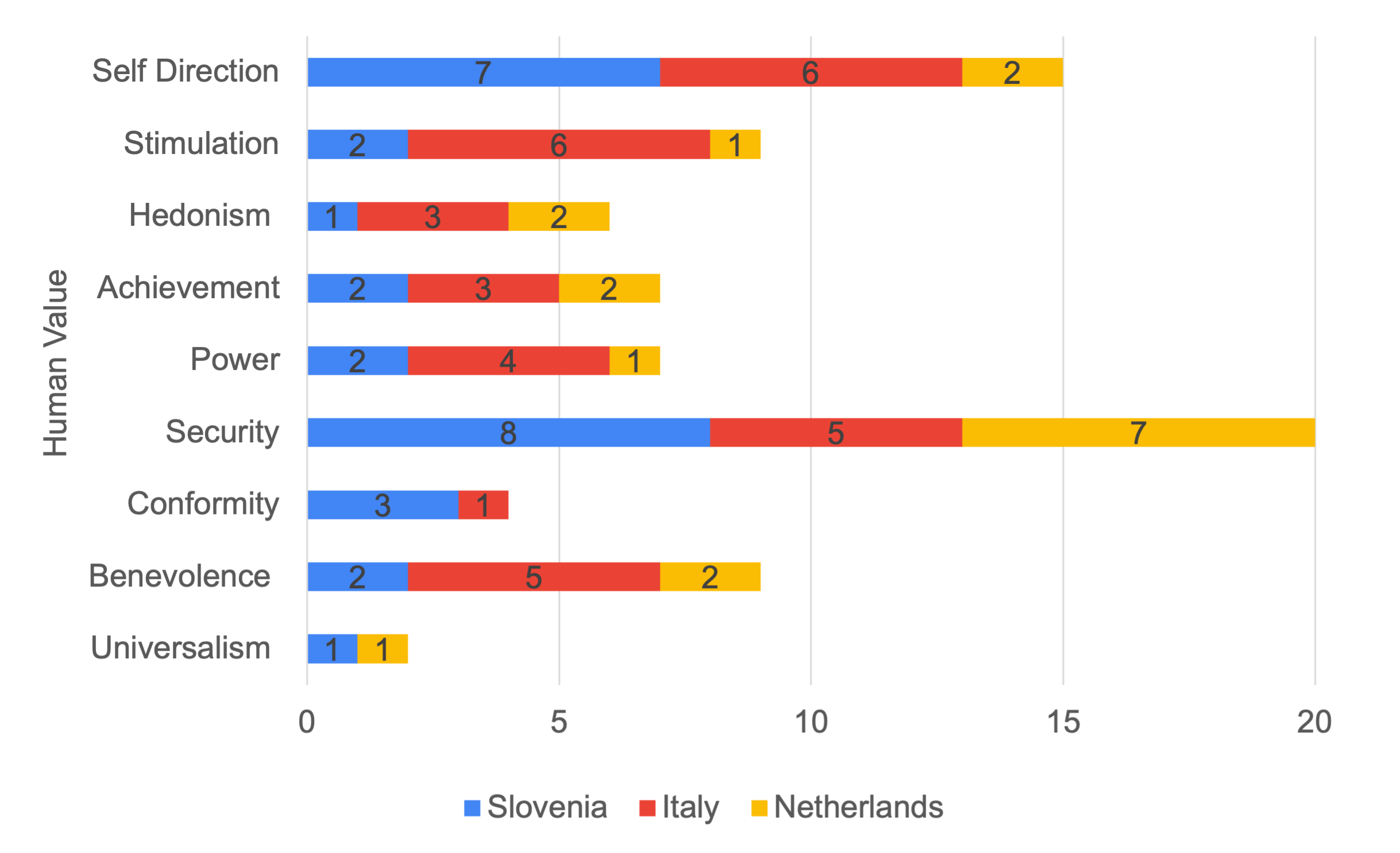}
   \caption{Presence of human values in pilots' motivational goal models}
\label{distribution_of_human_values_w.r.t_pilots}
\vspace{-6mm}
\end{figure}

The presence of human values for each pilot is depicted in Fig. \ref{distribution_of_human_values_w.r.t_pilots}. Out of 9 human values, 7 were shared by all three pilots. The value categories Universalism and Conformity could not be found in the goal models of the Italian and Dutch pilots, respectively. The most prominent human value category for the Slovenian and Dutch pilots is Security. In contrast, the top value categories for the Italian pilot are Self-direction and Stimulation.

Although all pilots are parts of the same Pharaon project in a similar healthcare and well-being domain, results still vary for each pilot, which can likely be attributed to different functional goals, roles and emotional goals. As Italian and Slovenian pilots shared similar goals, which can be seen in Fig. \ref{Frequencies_of_emotional goal_ per_pilot}, human values discovered for these two pilots are comparable. We can conclude that in addition to emotional goals, roles and functional goals also impact the presence of human values. Therefore, the presence of human value in different pilots can not be generalized.

%% file: Tables/Italian_human_value.tex
\begin{table*}
\centering
\small

\resizebox{\textwidth}{!}{%
\begin{tabular}{|p{7cm}|p{1cm}|p{2cm}|p{8cm}|}
\hline
\textbf{Functional goal} &
  \textbf{Role} &
  \textbf{Emotional Goals} &
  \textbf{Human Values} \\\hline
Manage health &
  OA &
  Informed &
  \begin{tabular}[c]{@{}l@{}}Self-direction  : Independence, Self-respect    \\ Security : Healthy, Sense of belonging\end{tabular} \\\hline
Manage health &
  ICG &
  Informed &
  \begin{tabular}[c]{@{}l@{}}Power : Social power \\ Benevolence : Helpful, Responsible\end{tabular} \\ \hline
Manage health &
  OA &
  Empowered &
  \begin{tabular}[c]{@{}l@{}}Self-direction  : Freedom, Choosing own goals \\ Achievement: Capable\end{tabular} \\ \hline
Manage health &
  OA &
  In touch &
  \begin{tabular}[c]{@{}l@{}}Security : Sense of belonging\\ Benevolence : True friendship\end{tabular} \\\hline
Manage health &
  OA &
  Involved &
  Self-direction  : Choosing own goals, Independence \\ \hline
Manage health &
  PCG &
  Aware &
  \begin{tabular}[c]{@{}l@{}}Achievement : Influential, Capable\\ Conformity : Honoring parents and Elders        \\ Power : Authority\end{tabular} \\ \hline
Manage health &
  OA &
  Stress free &
  \begin{tabular}[c]{@{}l@{}}Hedonism : Pleasure, Enjoying life \\ Self-direction : Independence, Privacy\end{tabular} \\ \hline
\begin{tabular}[c]{@{}l@{}}Contact relatives, \\ Make behavior suggestions\end{tabular} &
  ICG &
  Reassured &
  \begin{tabular}[c]{@{}l@{}}Benevolence  : Responsible, Helpful \\ Power : Recognition, Authority\end{tabular} \\ \hline
Make behavior suggestions &
  OA &
  Safe &
  Security : Healthy \\ \hline
Stimulation &
  OA &
  Active &
  Stimulation : Varied life, Excitement in Life \\ \hline
Provide feedback on the exercise suggestion (physical) &
  OA &
  Motivated &
  \begin{tabular}[c]{@{}l@{}}Stimulation : Excitement in Life \\ Achievement : Ambitious\end{tabular} \\ \hline
Provide feedback on the exercise suggestion (cognitive) &
  OA &
  Motivated &
  \begin{tabular}[c]{@{}l@{}}Stimulation : Excitement in Life  \\ Hedonism : Pleasure, Enjoying life\end{tabular} \\ \hline
Socialize &
  OA &
  Empowered &
  \begin{tabular}[c]{@{}l@{}}Self-direction : Freedom, Choosing own goals\\ Power : Recognition, Preserving my own image\end{tabular} \\ \hline
Maintain relationships &
  OA &
  Involved &
  \begin{tabular}[c]{@{}l@{}}Benevolence : True friendship\\ Stimulation : Excitement in life\\ Security : Sense of belonging\end{tabular} \\ \hline
Maintain relationships &
  OA &
  In touch &
  \begin{tabular}[c]{@{}l@{}}Benevolence : True friendship\\ Stimulation : Excitement in life\\ Security : Sense of belonging \end{tabular} \\ \hline
  
\end{tabular}%
}

\vspace{3mm}
OA= Older Adult, ICG= Informal Caregiver, PCG= Professional Caregiver
\caption{Mapping of goal models of the Italian pilot to the human values by Schwartz\\}
\label{tab:italian_human_value}
\vspace{-4mm}

\end{table*}

%% file: Disscussion.tex
\section{Discussion}\label{sec:Discussion}
Overall, we can conclude from our study that it is possible to identify human values in motivational goal models. We identified human values according to the theory of human values by Schwartz in motivational goal models that were created for representing early user requirements in an EU project on healthcare and well-being for older adults and their caregivers. 
In summary, we answer the research questions (RQs) that were posed in 
Section \ref{sec:intro} as follows:
\begin{itemize}
    \item \textbf{RQ1:} It is possible to identify human values in motivational goal models. For this, we propose mapping emotional goals from motivational goal models to the human values by Schwartz, considering also the context provided by roles and functional goals. In our industrial case study, we identified in six goal models from three pilots of the Pharaon project human values belonging to 9 out of 10 human value categories and 28 out of 58 value items based on the Schwartz theory of human values.
    \item \textbf{RQ2:} We found Security and Self-direction are the most prominent human values in the analysed three pilots, with the corresponding occurrence frequencies 20 and 15. The common prevalent human value category for all three pilots is Security. All three pilots share 7 out of 9 identified value categories, but these values are not evenly distributed, which can likely be attributed to different goal models.
\end{itemize}
In the following, we will discuss the answers to the RQs in more detail.

With respect to the research question RQ1, on the possibility of identification of human values in motivational goal models, we collected data from an ongoing EU project. We considered three pilots -- Italy, Slovenia, and the Netherlands -- and analysed their motivational goal models. Based on our results, we propose mapping emotional goals from goal models to human values based on the theory of human values by Schwartz, in the context of the corresponding functional goals and roles. We performed qualitative analysis described in Section \ref{sec:Data Analysis} and demonstrated that 9 out of 10 human values could be mapped from emotional goals in the context of roles and functional goals. Our results did not find the human value Universalism for any of the analysed pilots. This can be due to the nature of the project.

We observed during the analysis that different human values could be identified for the same emotional goal. A possible reason for this could be functional goals or roles in the models providing different context for the emotional goal. For example, in Table \ref{tab:italian_human_value}, the emotional goal Informed elicited for the respective roles Older Adult and Informal Caregiver is associated with the identical functional goal Manage health. However, the human values identified for Informed differ in these two cases because the roles associated with each case are different: Older Adult vs Informal Caregiver (i.e., relative or friend). According to the corresponding scenario, an older adult wants to feel informed about his/her health management. Therefore, the identified human values are Self-direction and Security. Differently, when an informal caregiver expects to be informed about managing the health of an older adult she/he is taking care of, the identified human values are Power and Benevolence. The explanation is that if a caregiver is informed about the health of the older adult, he/she can be helpful and responsible, but for giving good care, he/she also needs to have power over people and resources. Similarly, the emotional goal Empowered is related to the same role Older Adult, but has been mapped to partially different human values due to the different functional goals Manage health and Socialize. Therefore, we can conclude that emotional goals solely do not determine human values, and functional goals and roles influence the results, defining the context for the emotional goals. 


With respect to the research question RQ2, we analyzed human values by value category and items shown in Table I. The results of our analysis demonstrate that almost 45\% out of 79 value occurrences from 9 distinct value categories belong to the value categories Security and Self-direction with the respective occurrence frequencies 20 and 15. The least apparent values in our results are Universalism and Conformity, with the respective occurrence frequencies 2 and 4. We found the value items Sense of belonging and Healthy as the most frequent value items. Since our industrial case study is concerned with improving the health and quality of life of older adults, the most frequent role appearing in goal models is Older Adult. The most frequent value items and role indirectly validate our results, because improving sense of belonging to a community and health of older adults are among the main objectives of the Pharaon project. 

The granularity of value items included by the Schwartz theory of human values can be utilized to better inform the pilots and practitioners of the Pharaon project. For example, Security is the most prevalent human value identified in the analysed pilots, where the most common value items are Sense of belonging and Healthy. This information can be helpful for developers and practitioners when making design and development decisions. For example, they can consider human values when creating user stories to improve user experience and acceptance by main end users of the Pharaon ecosystem -- older adults -- and their caregivers. The identified and embedded human values should also be validated with end users, which we were not able to do yet because of the current stage and large scope of the Pharaon project. Overall, all three pilots share 7 out of 9 human values, but these values are not evenly distributed due to different goal models which have been created by different pilots. Therefore, the presence of human value across different pilots can not be generalized.

Moreover, our findings can and should be extended to developing other similar platforms of digital healthcare and well-being services for older adults and the related stakeholder groups. Our findings indicate that the usage of motivational goal modelling for eliciting and representing early requirements enables to consider from the very start of the developing process human values important for older adults and other stakeholders, such as Security and Self-direction.


\subsection{Threats to Validity}
A threat to construct validity arises from the decision to use the Schwartz theory of human values \cite{schwartz2012overview}. There are also other theories of human values \cite{rokeach1973nature, hofstede1984hofstede,gouveia2014functional} that could be used in place of the Schwartz theory. However, none of these theories has been applied in software engineering, while the Schwartz theory has already been widely used in software engineering \cite{ferrario2016values, whittle2019case, nurwidyantoro2022human, hussain2022can} and requirements engineering \cite{proynova2011investigating, alatawi2018psychologically, perera2020continual, obie2021first}. Another possible threat to construct validity is the definition of human values; they might have been vague for the analysts. We mitigated this threat by referring to the literature on the Schwartz theory of human values and its applications in software engineering and requirements engineering during each analysis session. This helped to create a basic understanding of the values' theory. Moreover, we undertook detailed discussion sessions that were moderated by a social scientist who is an expert in the Schwartz theory of human values, thus helping in cases that were initially missing the social science perspective.

Another threat to internal validity involves the analysis process, as mapping emotional goals to human values can be subjective and error-prone. For mitigation, we performed an iterative analysis process, consisting of the trial session, mock session, and individual analysis sessions. Furthermore, we undertook a detailed discussion session between the analysts and moderator, who is an expert in the Schwartz theory of human values, to resolve disagreements and conflicts, and finalized the human values only after reaching a mutual agreement. This process reduced the number of errors due to misinterpretations.

A threat to external validity is the generalizability of our results due to the Pharaon project that we chose for our study. Our project is focused on the domains of healthcare and well-being and is limited by the size of the data available to us. We mitigated this threat by analyzing three pilots from three different countries. However, we acknowledge that it may not be possible to generalize our findings to other domains or even to a similar domain. More research needs to be conducted to explore this. We have provided the dataset that includes the goal models and project report for the study's replicability. Although, human value is subjective, which can cause different results.  To this end, if other researchers follow the process we provided in Fig \ref{process}, they should get comparable results.


%% file: Relatedwork.tex
\section{Related Work}
\label{sec:Related}

The foundation for representing human values through emotional requirements is discussed in \cite{sterling2009art}, where emotional requirements are represented as quality goals and further elaborated in \cite{miller2014requirements}, where the agent paradigm was introduced for requirements elicitation and representation. Emotional requirements were explicitly introduced and exemplified by the case study of emergency alarm systems for older adults in \cite{miller2015emotion}. The same case study was extended for elaborating emotion-oriented requirements engineering in \cite{curumsing2019emotion}, based on \cite{sterling2009art} and \cite{miller2015emotion}. The relevance of emotional requirements in healthcare was demonstrated by two case studies in \cite{taveter2019method}. However, none of these studies applies any existing theory of human values to map the emotional goals to the values by multiple stakeholder groups.

The study \cite{zdravkovic2015capturing} provides a framework for capturing consumer preferences in terms of human values and mapping them to soft-goals of the Tropos methodology \cite{giorgini2005goal}. The article \cite{singh2022modelling} proposes to extend soft-goals of the Tropos methodology with \textit{contexts} representing ``human-centric aspects of end-users'' that are associated with tasks contributing to achieving the soft-goals. The work \cite{perera2020continual} presents a method for reasoning about human values in software systems by means of goal models of the Tropos methodology where goals represent human values and tasks -- features of software systems. The paper \cite{alatawi2018psychologically} puts forward a method and notation for representing personal values and relating them to motivations modelled as quality goals and emotions captured as emotional goals of motivational goal models. The work \cite{hassettfeel} proposes to link organisational values like \textit{gender-inclusiveness} with personal human values through emotional goals of motivational goal models, such as Valued, Supported, and Inclusive. As compared to the mentioned studies, our approach described in this paper stands out because (i) it is simple as human values are mapped from emotional goals in the context of roles and functional goals; (ii) is has a strong theoretical foundation, as the approach is based on the theory of constructed emotion \cite{iqbal2022theory,taveter2019method}, in addition to the Schwartz theory of human values; (iii) it lends itself more easily into practical applications because motivational goal models can be transformed in a straightforward manner into detailed requirements in the form of user stories \cite{oliveira2021transitioning,tenso2017enhancing}, as has been demonstrated in the Pharaon project.


%% file: Conclusion.tex
\section{Conclusions}
\label{sec:conclusion}
Our study demonstrates that human values can be identified in motivational goal models, which is one of the artefacts in goal-based requirements engineering. From six goal models of three pilots of the Pharaon project -- Italy, the Netherlands, and Slovenia -- 
we identified human values belonging to 9 out of 10 value categories and 28 out of 58 value items based on the Schwartz theory of human values. All in all, we identified 79 value occurrences from 9 distinct value categories by Schwartz. We found that almost 45\% of all the value occurrences belonged to the value categories Security and Self-direction. All three pilots shared 7 out of 9 identified value categories. We also identified the value items Sense of belonging and Healthy as the most frequent value items, both of which belong to the value category Security. We also found that analysing the emotional goals solely is not sufficient for identifying human values, as also the broader context in the form of functional goals and roles should be used. 

The relevance of our work does not lie in identifying human values in motivational goal models \textit{per se} bur rather paves the way for including human values in goal-based requirements so that they could be elaborated into detailed requirements and software systems preserving human values embedded in them. The corresponding recommendation for practitioners arising from our work is to use motivational modelling to represent early user requirements, because, as we have shown in this paper, emotional goals included in motivational goal models can be mapped to human values, considering the context provided by the roles and functional goals. Our results indicate that any requirements engineer working on the project at hand who has familiarized herself with the Schwartz theory of human values is able to perform this kind of mapping. Early requirements represented by motivational goal models can be elaborated into scenarios and user stories, preserving the human values captured by motivational goal models. We will demonstrate a methodology for doing this as an important part of our future work.

Another important area of our future work will be to include quality (i.e., non-functional) goals in the analysis of human values. To support the inclusion of human values in the requirements of sociotechnical systems, we also plan to create a ``dictionary'' mapping emotional goals found in the research literature into categories of human values associated with them.

In the future, we will also extend our work to the refined theory of human values by Schwartz \cite{schwartz2012refining} and to the remaining three pilots of the Pharaon project and other similar projects.

